\documentstyle[aps,preprint,eqsecnum,prd]{revtex}

\newbox\rotbox

\begin{document}
% The following 3 lines are for preprints only.
%\noindent{\it Submitted to:} \hfill {OSU-PP \#94-0333}
%
%\noindent{\it Phys. Rev. D}  \hfill
%
%\vspace{12pt}
%\def\gtwid{\mathrel{\raise.3ex\hbox{$>$\kern-.75em
%     \lower1ex\hbox{$\sim$}}}}
%\def\ltwid{\mathrel{\raise.3ex\hbox{$<$\kern-.75em
%     \lower1ex\hbox{$\sim$}}}}

\draft

\title{Nucleon properties from unconventional interpolating fields}

\author{Derek B. Leinweber\cite{presaddr}}

\address{
Department of Physics, The Ohio State University \\ 174 West 18'th
Avenue, Columbus, OH 43210--1106 \\ }

\date{April 1994}

\maketitle

\begin{abstract}
Interpolating fields, used to excite hadrons from the QCD vacuum in
nonperturbative field-theoretic investigations of hadron properties,
are explored with an emphasis on unconventional nucleon interpolators.
The QCD continuum model for excited state contributions to QCD
correlation functions is a central element in extracting the physics
contained in these alternate correlation functions.  The analysis
confirms the independence of nucleon properties obtained from
different interpolating fields.  However, this independence comes
about in a trivial manner.  These results provide a resolution to the
long standing debate over the optimal nucleon interpolating field to
be used in QCD Sum Rule analyses.
\end{abstract}


\narrowtext

\section{INTRODUCTION}
\label{intro}

   One of the instrumental operators of nonperturbative
field-theoretic investigations of hadron structure is the hadron
interpolating field. This operator is used to excite a hadron of
specified quantum numbers from the QCD vacuum.  It has long been
established that there are two independent interpolating fields with
no derivatives having the quantum numbers of spin 1/2 and isospin 1/2.
Both are expected to excite the ground state nucleon from the vacuum.
Various linear combinations of these interpolators are used in
nonperturbative approaches to QCD.  What distinguishes the different
approaches is the manner in which the propagation of quarks in the
QCD vacuum is determined.

   Numerical simulations of the theory via lattice regulation is the
only method for probing deep into the nonperturbative regime of QCD.
Exploitation of the Operator Product Expansion (OPE) in QCD Sum Rules
(QCD-SR) allows the near perturbative regime of QCD to be explored.
Modeling of the QCD vacuum via instanton fluctuations in the Random
Instanton Liquid Model (RILM) has also produced some new insights into
QCD.  While there are formal field theoretic arguments indicating
nucleon properties are independent of the interpolating field, a
demonstration of this in practice is an important test of these
approaches to nonperturbative field theory.

   Some attention has been given previously to alternate nucleon
interpolators on the lattice \cite{leinweber91,bowler84}.  In these
analyses conclusions were limited as the correlation functions
deteriorated too quickly for ground state properties to be determined.
In a previous paper \cite{leinweber94a} it was established that the
properties of the lowest lying state may be extracted from the first
few points of two-point correlation functions with the use of a pole
plus QCD continuum model inspired by QCD Sum Rule analyses.  In this
paper, these techniques are used to investigate nucleon properties
obtained from correlation functions of unconventional nucleon
interpolating fields.  Some of these correlation functions suffer a
loss of signal prior to a clear ground state domination and the QCD
continuum model becomes a central element in fitting the correlation
functions.

   In the QCD-SR field there is a history of argument over the optimum
nucleon interpolating field to be used in analyses
\cite{chung82,chung84,dosch89a,ioffe81,ioffe83,leinweber90,furnstahl92,jin94}.
This issue is recognized to be of paramount importance and remains
unsettled \cite{shifman92}.  While some advocate an interpolating
field for which the leading terms of the OPE are stationary with
respect to the interpolating field mixing parameter
\cite{chung84,dosch89a}, others argue that a balance between
OPE convergence\footnote{Here and in the following, ``convergence'' of
the OPE simply means that the highest dimension terms considered in
the OPE, with their Wilson coefficients calculated to leading order in
perturbation theory, are small relative to the leading terms of the
OPE.} and QCD continuum contributions must be maintained
\cite{ioffe83,leinweber90,furnstahl92,jin94}.

   Ideally, one would like to simply calculate with alternate
interpolating fields and confirm that the nucleon properties remain
unchanged.  However, the limitations of the QCD-SR approach have
prevented one from doing this in practice.  Limitations include
uncertainties in the values of lower dimension condensates,
factorization of higher dimension operators, OPE truncation and
convergence issues, uncertainties surrounding the role of direct
instanton contributions to the sum rules and uncertainties in the
reliability of the continuum model for excited states.  Fortunately,
the lattice approach is not plagued with the same limitations and the
following analysis resolves this long standing debate.

   The format of this paper is as follows.  Section \ref{lattcorrfun}
introduces the interpolating fields explored in this analysis, the
lattice techniques used, and issues encountered in relating lattice
and continuum (lattice spacing $a \to 0$) formalisms.  Section
\ref{contmodel} highlights the QCD continuum model derivation.  The
analysis of the correlation functions is presented in Section
\ref{lattcorrfits} for each interpolating field combination.  The
results are compared with other approaches to QCD in Section
\ref{compare}.  Finally the conclusions regarding interpolating field
invariance and the optimal nucleon interpolator for QCD Sum Rules are
summarized in Section
\ref{conclusions}.

\section{LATTICE CORRELATION FUNCTIONS}
\label{lattcorrfun}

\subsection{Interpolating Fields}

   The commonly used interpolating field for the proton in lattice
calculations has the form
\begin{equation}
\chi_1(x) = \epsilon^{abc}
                 \left ( u^{Ta}(x) C \gamma_5 d^b(x) \right ) u^c(x)
\, .
\label{chiN1}
\end{equation}
Here, we follow the notation of Sakurai \cite{sakurai67}.  $C =
\gamma_4 \gamma_2$ is the charge conjugation matrix, $a,\ b,\ c$ are
color indices, $u(x)$ is a $u$-quark field, and the superscript $T$
denotes transpose.  Dirac indices have been suppressed.

   In the sum rule approach, it is common to find linear combinations
of this interpolating field and
\begin{equation}
\chi_2(x) = \epsilon^{abc}
                 \left ( u^{Ta}(x) C d^b(x) \right ) \gamma_5 u^c(x)
\, ,
\label{chiN2}
\end{equation}
which vanishes in the nonrelativistic limit.  {\it A priori}, there is
no reason to exclude such an interpolating field \cite{chung82}, as
the quark field operators of (\ref{chiN2}) annihilate the light
current quarks of QCD.  Of course, these quarks are highly
relativistic when bound in the nucleon.  With the use of the Fierz
relations, the combination of the above two interpolating fields with
a relative minus sign may be written
\begin{eqnarray}
\chi_{\rm SR}(x) &=& \epsilon^{abc}
                 \left ( u^{Ta}(x) C \gamma_\mu u^b(x) \right )
                 \gamma_5 \gamma^\mu d^c(x) \nonumber \\
           &=& 2 \left ( \chi_2 - \chi_1 \right ) \, ,
\label{chiSR}
\end{eqnarray}
giving the proton interpolating field often found in sum rule
calculations \cite{ioffe83,leinweber90,furnstahl92,jin94}.  The
alternate QCD Sum Rule interpolating field is
\begin{eqnarray}
\chi_{\rm A}(x) &=& {1 \over 2} \epsilon^{abc}
                 \left ( u^{Ta}(x) C \sigma_{\mu \nu} u^b(x) \right )
                 \sigma^{\mu \nu} \gamma_5 d^c(x) \nonumber \\
           &=& 2 \left ( \chi_2 + \chi_1 \right ) \, .
\label{chiA}
\end{eqnarray}

   In this analysis we will consider both interpolating fields
introduced in (\ref{chiN1}) and (\ref{chiN2}) and their interference
terms such that any linear combination of these interpolating fields
may be investigated.

\subsection{Correlation Functions at the Quark Level}

   Hadron masses are determined through the consideration of two-point
correlation functions.  Here we consider the nucleon correlator
\begin{equation}
G_2(t,\vec p) = \sum_{\vec x}\, e^{-i \vec p \cdot \vec x}
{\rm tr} \Bigl [ \Gamma_4
\bigm < 0 \bigm | T \{ \chi_N (x)\, \overline \chi_N (0) \} \bigm |
0 \bigm > \Bigr ]
\label{twopoint}
\end{equation}
where $\chi_N(x)$ may be either (\ref{chiN1}) or (\ref{chiN2}),
$\Gamma_4 = (1 + \gamma_4)/4$ projects positive parity states for
$\vec p = 0$, and tr indicates the trace over Dirac indices.

   Correlation functions at the quark level are obtained through the
standard procedure of contracting out time-ordered pairs of quark
field operators.  For the octet baryons it is convenient to define the
correlation function
\begin{eqnarray}
\lefteqn{
{\cal F}(S_{f_1},S_{f_2},S_{f_3}) = } \hspace{2.2cm} \nonumber \\
\epsilon^{abc} \epsilon^{a'b'c'}
\biggl \{
&& S_{f_1}^{a a'}(x,0) \, tr \left [ S_{f_2}^{b b'}(x,0) \,
   S_{f_3}^{c c' \, T}(x,0)
   \right ] \nonumber \\
&& + S_{f_1}^{a a'}(x,0) \, S_{f_3}^{c c' \, T}(x,0) \,
   S_{f_2}^{b b'}(x,0) \biggr \} \, ,
\label{calFdef}
\end{eqnarray}
where $S^{a a'}(x,0) = T \left \{ q^a(x), \, \overline q^{a'}(0)
\right \}$ and $f_1$, $f_2$, $f_3$ are flavor labels.  For the proton
interpolating field $\chi_1$ of (\ref{chiN1}), the two-point function
may be written
\begin{equation}
G_2(t,\vec p) = \sum_{\vec x} e^{-i \vec p \cdot \vec x}
   tr \left [ \Gamma_4 \, \,
   {\cal F} \left ( S_u, \, S_u, \, \widetilde C S_d
   \widetilde C^{-1} \right ) \right ] \, ,
\label{latticeCorrel1}
\end{equation}
where $\widetilde C = C \gamma_5$. Similarly the two-point function
corresponding to $\chi_2$ of (\ref{chiN2}) may be written in the form
\widetext
\begin{equation}
G_2(t,\vec p) = \sum_{\vec x} e^{-i \vec p \cdot \vec x}
   tr \left [ \Gamma_4 \, \,
   {\cal F} \left ( \gamma_5 S_u \gamma_5, \,
   \gamma_5 S_u \gamma_5, \, \widetilde C S_d
   \widetilde C^{-1} \right ) \right ] \, .
\label{latticeCorrel2}
\end{equation}
The interference contributions of these two interpolating fields are
\begin{equation}
G_2(t,\vec p) =  \sum_{\vec x} e^{-i \vec p \cdot \vec x}
   tr \biggl [ - \Gamma_4 \, \, \biggl \{
   {\cal F} \left ( S_u \gamma_5,\,
   S_u \gamma_5, \, \widetilde C S_d \widetilde C^{-1} \right )
   + {\cal F} \left ( \gamma_5 S_u,\, \gamma_5 S_u, \,
   \widetilde C S_d \widetilde C^{-1}
   \right ) \biggr \} \biggr ] \, ,
\label{latticeCorrelI}
\end{equation}

\narrowtext

\subsection{Lattice Techniques}

   Here we briefly summarize the lattice techniques used in the
following calculations.  Additional details may be found in Ref.\
\cite{leinweber91}.  Wilson's formulation is used for both the gauge
and fermionic action $(r=1)$.  $SU(2)$-isospin symmetry is enforced by
equating the Wilson hopping parameters $\kappa_u = \kappa_d = \kappa$.
Three values of $\kappa$ are selected and are denoted $\kappa_1 =
0.152$, $\kappa_2 = 0.154$ and $\kappa_3 = 0.156$.  To make contact
with the physical world, the mass and interpolating field coupling
strengths calculated at the three values of $\kappa$ are linearly
extrapolated to $\kappa_{\rm cr}=0.159\,8(2)$ where an extrapolation
of the squared pion mass vanishes.  Differences between linear
extrapolations to $m_\pi=0$ as opposed to the physical pion mass are
small and are neglected in the following.

  Twenty-eight quenched gauge configurations were generated by the
Cabibbo-Marinari \cite{cabibbo82} pseudo-heat-bath method on a $24
\times 12 \times 12 \times 24$ periodic lattice at $\beta=5.9$.
Configurations were selected after 5000 thermalization
sweeps from a cold start, and every 1000 sweeps thereafter
\cite{correlations}.

   Dirichlet boundary conditions are used for fermions in the time
direction.  Time slices are labeled from 1 to 24, with the
$\delta$-function source at $t=4$.  To minimize noise in the Green
functions, the parity symmetry of the correlation functions, and the
equal weighting of $\{U\}$ and $\{U^*\}$ gauge configurations in the
lattice action are exploited.  The nucleon mass determined from
$\chi_1$ of (\ref{chiN1}) is used to set the lattice spacing.  This
estimate lies between other estimates based on the string tension or
the $\rho$-meson mass.  The lattice spacing is determined to be
$a=0.132(4)$ fm and $a^{-1} = 1.49(5)$ GeV.

   Statistical uncertainties in the lattice correlation functions are
estimated by a single elimination jackknife \cite{efron79}.  A
covariance matrix fit of the pole plus QCD continuum model over a
range of 7 or more time slices is likely to be unreliable for 28 gauge
configurations \cite{michael94}.  Instead we use standard statistical
error analysis in which correlations among the fit parameters are
accounted for.  The Gauss-Newton method is used to minimize $\chi^2$.
Uncertainties are taken from the standard error ellipse \cite{rpp92}
at $\chi^2 = \chi^2_{\rm min} + 1$.

\subsection{Operator Mixing}

   The implementation of Wilson fermions on the lattice induces mixing
between the composite nucleon interpolating fields \cite{richards87}
of (\ref{chiN1}) and (\ref{chiN2}), reflecting the breaking of chiral
symmetry.  In Ref. \cite{richards87} the mixing is argued to occur
between
\begin{mathletters}
\begin{eqnarray}
O_\alpha &\equiv& \epsilon^{abc}
                  \left ( u^{Ta}_R(x) C d^b_R(x) \right ) u^c_L(x)
\, , \\
&=& {1 \over 4} \left ( 1 - \gamma_5 \right )
                \left ( \chi_1 - \chi_2 \right ) \, , \nonumber \\
O_\beta &\equiv& \epsilon^{abc}
                  \left ( u^{Ta}_L(x) C d^b_L(x) \right ) u^c_L(x)
\, , \\
&=& -{1 \over 4} \left ( 1 - \gamma_5 \right )
                 \left ( \chi_1 + \chi_2 \right ) \, , \nonumber
\end{eqnarray}
and a third operator
\begin{equation}
O_\gamma  \equiv \epsilon^{abc}
          \left ( u^{Ta}(x) C \gamma_\rho \gamma_5 d^b(x) \right )
          \gamma_L \gamma^\rho u^c(x) \, .
\end{equation}
\end{mathletters}
Here
\begin{mathletters}
\begin{equation}
u_R = {1 \over 2} \left ( 1 + \gamma_5 \right ) u \, , \qquad
u_L = {1 \over 2} \left ( 1 - \gamma_5 \right ) u \, ,
\end{equation}
\begin{equation}
\gamma_R = {1 \over 2} \left ( 1 + \gamma_5 \right ) \, ,
\quad {\rm and} \quad
\gamma_L = {1 \over 2} \left ( 1 - \gamma_5 \right ) \, .
\end{equation}
\end{mathletters}
However, there are only two operators having isospin-1/2 and spin-1/2
and it is possible to demonstrate
\begin{equation}
O_\gamma = - 2 O_\alpha \, ,
\end{equation}
via Fierz transformations.  For $\chi_1$ and $\chi_2$ the expressions
up to one-loop order in perturbation theory relating the operator
matrix elements in the Pauli-Villars (PV) and Lattice (L) schemes are
\cite{richards87}
\mediumtext
\begin{eqnarray}
\chi_1^{\rm PV} &=& \chi_1^{\rm L}
      - {\alpha_s \over 4 \pi} \,
        \left [ -2 \, \ln Q^2 a^2 +
        \left ( C_1^{\rm L} - C_2^{\rm L} - 2 C_3^{\rm L} \right )
        \right ] \, \chi_1^{\rm L}
      - {\alpha_s \over 4 \pi} \, 2 C_3^{\rm L} \, \chi_2^{\rm L}
\, , \\
\chi_2^{\rm PV} &=& \chi_2^{\rm L}
                  - {\alpha_s \over 4 \pi} \,
                    \left [ -2 \, \ln Q^2 a^2 +
                    \left ( C_1^{\rm L} + C_2^{\rm L} \right )
                    \right ] \, \chi_2^{\rm L} \, ,
\end{eqnarray}
\narrowtext
where
\begin{equation}
C_1^{\rm L} = 37.91, \quad C_2^{\rm L} = -3.21, \quad {\rm and} \quad
C_3^{\rm L} = -0.80,
\end{equation}
for the Wilson parameter $r=1$.  The important point is that the
interpolating field $\chi_2$ does not mix with $\chi_1$ to one-loop
order.  Moreover, the mixing of $\chi_1$ with $\chi_2$ is negligible.
Hence, it is possible to identify the properties of these
interpolating fields determined on the lattice with those of their
continuum $(a \to 0)$ counterparts to a good approximation.

   The dominant contribution to the coefficient $C_1^{\rm L}$ in the
above expressions is from the self-energy corrections to the quark
external lines.  These corrections are accounted for in the mean-field
improved approach, and the remaining renormalization $Z_\chi$
associated with composite operators is relatively small.  The
principle renormalization constant $C_1^{\rm L}$ has been determined
in the mean-field approach \cite{lepage93} and is used in the
following.  The nucleon coupling strength, $\lambda_N$, is determined
in absolute terms, without resorting to a ratio of the QCD continuum
contributions as done in \cite{chu93a,chu93b}.  In particular, the
renormalization at the scale of $1/a$ is
\begin{equation}
\chi^{\rm Continuum} = {Z_{\chi_N} \over a^{9/2}} \,
\chi^{\rm Lattice} \,
\left ( 1 - { 3 \kappa \over 4 \kappa_{\rm cr}} \right )^{3/2} \, ,
\end{equation}
and $Z_{\chi_N} = (1 - 0.73 \, \alpha_V) \simeq 0.80$ at $\beta =
5.9$.
%
%\alpha_V = 0.2763  Z_\chi = 0.80
The $\kappa$ dependence of this wave function renormalization is very
different from the na\"{\i}ve normalization
\begin{equation}
\chi^{\rm Continuum} = {1 \over a^{9/2}} \, \chi^{\rm Lattice} \,
\left ( 2 \kappa \right )^{3/2} \, ,
\end{equation}
and is crucial to recovering the correct mass independence of the
Wilson coefficient of the identity operator.

\section{THE QCD CONTINUUM MODEL}
\label{contmodel}

   Here we briefly review the QCD continuum model implementation in
Euclidean space as examined in detail in Ref.\ \cite{leinweber94a}.
We start with the two-point correlation function of (\ref{twopoint}).
At the phenomenological level, one inserts a complete set of states
$N^i$ and defines
\begin{equation}
\bigm < 0 \bigm | \chi_N (0) \bigm | N^i,p,s \bigm > \, =
\lambda_N^i\, u(p,s) \, ,
\end{equation}
where the coupling strength, $\lambda_N^i$, measures the ability of
the interpolating field $\chi_N$ to annihilate the i'th excitation
with nucleon quantum numbers.  For $\vec p = 0$ and Euclidean time $t
\to \infty$, the ground state dominates and $G_2(t) \to \lambda_N^2
e^{-M_N t}$.  The spectral representation is defined by
\begin{equation}
G_2(t) = \int_0^\infty \, \rho(s) \, e^{-st} \, ds \, ,
\label{SpectralRepr}
\end{equation}
and the spectral density is,
\begin{equation}
\rho(s) = \lambda_N^2 \, \delta(s-M_N) + \zeta(s)
\end{equation}
where $\zeta(s)$ provides the excited state contributions.

   The form of the spectral density used in the QCD continuum model is
determined by the leading terms of the OPE surviving in the limit $t
\to 0$.  Here, the combination of interpolators $\chi_1 \overline
\chi_1$ is considered.  The derivation of the QCD continuum model
contributions to other correlation functions proceeds in an analogous
fashion.  In Euclidean space, $G_2(t)$ has the following OPE
\begin{eqnarray}
G_2(t) &\simeq& {3 \cdot 5^2 \over 2^8 \pi^4} \, \biggl ( \,
{1 \over t^6} + {28 \over 25} \, {m_q a \over t^5}
              + {14 \over 25} \, {m_q^2 a^2 \over t^4} \nonumber \\
            &&- {56 \pi^2 \over 75} \, {\bigm < : \overline q q :
                                        \bigm > a^3
                \over t^3}
              + \cdots \biggr ) \, .
\label{OPE11}
\end{eqnarray}

   The spectral density used in the QCD continuum model is defined by
equating (\ref{SpectralRepr}) and (\ref{OPE11}).  The QCD continuum
model is defined through the introduction of a threshold which marks
the effective onset of excited states in the spectral density.
Keeping the first two terms of (\ref{OPE11}), the phenomenology of
$G_2(t)$ is \widetext
\begin{mathletters}
\label{phenom}
\begin{eqnarray}
G_2(t) &=& \lambda_1^2 \, e^{-M_N t} + \xi \,
\int_{s_0}^\infty \, \rho(s) \, e^{-st} \, ds  \, ,
\label{phenomCrypt} \\
&=& \lambda_1^2 \, e^{-M_N t} + \xi \,
{3 \cdot 5^2 \over (2^8 \pi^4)} \,
e^{-s_0 t} \times \label{phenomExpl} \\
&& \quad \Biggl ( \left \{ {1 \over t^6} + {s_0 \over t^5}
+ {1 \over 2}\,{s_0^2 \over t^4} + {1 \over 6}\,{s_0^3 \over t^3}
+ {1 \over 24}\,{s_0^4 \over t^2} + {1 \over 120}\,{s_0^5 \over t}
\right \}
+ {28 \, m_q a \over 25} \,  \left \{
  {1 \over t^5} + {s_0 \over t^4}
+ {1 \over 2}\,{s_0^2 \over t^3} + {1 \over 6}\,{s_0^3 \over t^2}
+ {1 \over 24}\,{s_0^4 \over t}
\right \} \Biggr ) \, .  \nonumber
\end{eqnarray}
\end{mathletters}
\narrowtext
The parameter $\xi$ governs the strength of the QCD continuum model.
In the continuum limit $(a \to 0)$ $\xi = 1$ but here is optimized
with $\lambda_N$, $M_N$, and $s_0$ to account for enhancement of the
correlator in the short time regime reflecting lattice anisotropy.
$\xi$ is an overall QCD continuum model strength and is expected to be
independent of the quark mass.  With this approach, the effects of
lattice anisotropy may be absorbed through a combination of a larger
QCD continuum model strength $(\xi > 1)$ and marginally larger
threshold $(s_0)$.

   Infrared lattice artifacts are not a significant problem for this
approach as the Fourier transform weight $\exp(-i \vec p \cdot \vec
x)$ is correct for all propagator paths including those which wrap
around the lattice spatial dimensions.  The ultraviolet lattice cutoff
may be modeled in a manner similar to that for the QCD continuum
model.  However the modeling becomes insignificant by the second time
slice following the source.  Instead we simply discard the source and
first time slice when fitting the correlation functions.

\section{LATTICE CORRELATOR FITS}
\label{lattcorrfits}

\subsection{\pmb{$\bigm < \chi_1 \overline \chi_1 \bigm >$}
            Correlation Function }

   $\chi_1$ is the standard nucleon interpolating field used in
lattice analyses.  Its overlap with the nucleon ground state is
excellent.  This lattice correlation function is fit with
(\ref{phenom}) in a four parameter search of $\lambda_N$, $M_N$, $s_0$
and $\xi$ in analysis intervals from $t=6 \to t_f$ where $t_f$ ranges
{}from 11 through 23.  Figure \ref{NuclCorrFn} illustrates the lattice
data and these 13 fits at our intermediate value of quark mass.
Similar results are seen for $\kappa = 0.152$ and 0.156.

   The similarity of the 13 pole plus QCD continuum fits establishes
that the QCD continuum model effectively accounts for excited state
contaminations in the correlation functions and allows the extraction
of the ground state properties from a regime as small as $t=6 \to 11$.
For an in-depth examination of this correlator see Ref.\
\cite{leinweber94a}.  The quark mass dependence of $\lambda_N$ is
illustrated in Figure \ref{lambda1}.  Table \ref{Tab:chi11C5ope}
summarizes the fit parameters for the regime $t=6 \to 20$.

\subsection{\pmb{$\bigm < \chi_2 \overline \chi_2 \bigm >$}
            Correlation Function }

   Unlike QCD Sum Rule analyses, correlators may be determined on the
lattice for any interpolating field without regard to the restrictions
of OPE convergence issues or operator factorization assumptions.
In constructing the QCD continuum model, only the first few terms of
the OPE are required.  The OPE for the interpolating fields $\chi_2
\overline \chi_2$ is
\begin{eqnarray}
G_2(t) &=& {3 \cdot 5^2 \over 2^8 \pi^4} \,
\biggl ( \, {1 \over t^6} -
{4 \over 5} \, {m_q a \over t^5} - {2 \over 5} \, {m_q^2 a^2 \over
t^4} \nonumber \\ &&+ {8 \pi^2 \over 15} \, {\bigm < \overline q q
\bigm > a^3 \over t^3} + \cdots \biggr ) \, ,
\label{OPE22}
\end{eqnarray}
and the QCD continuum model is derived in an analogous manner to that
outlined in Section \ref{contmodel}.  The phenomenological side of
$G_2(t)$ is
\widetext
\begin{eqnarray}
G_2(t) &=& \lambda_N^2 \, e^{-M_N t} +
           \xi \, {3 \cdot 5^2 \over (2^8 \pi^4)} \,
           e^{-s_0 t} \, \times \nonumber \\
&& \Biggl (
\left \{ {1 \over t^6} + {s_0 \over t^5}
+ {1 \over 2}\,{s_0^2 \over t^4} + {1 \over 6}\,{s_0^3 \over t^3}
+ {1 \over 24}\,{s_0^4 \over t^2} + {1 \over 120}\,{s_0^5 \over t}
\right \}
- {4 \, m_q a \over 5} \,  \left \{
  {1 \over t^5} + {s_0 \over t^4}
+ {1 \over 2}\,{s_0^2 \over t^3} + {1 \over 6}\,{s_0^3 \over t^2}
+ {1 \over 24}\,{s_0^4 \over t}
\right \} \Biggr ) \, .  \label{phenomExpl2}
\end{eqnarray}
\narrowtext

   Figure \ref{chi22} illustrates the lattice correlation function and
the final fit.  The choice of $\Gamma_4$ in (\ref{twopoint}) projects
out positive parity nucleon states when $\vec p = 0$ and therefore the
correlation function must remain positive.  At $t=13$ the lattice
correlation function data changes sign, and indicates a loss of
signal.

   The fit from $t = 6 \to 12$ using a pole plus QCD continuum model
leads to fit parameters where the pole lies above the continuum
threshold.  The position of the pole is insignificant as its removal
has little effect on the $\chi^2$/dof.  Fixing the pole at the
previously determined nucleon masses returns an optimum value for
$\lambda_2^2$ which is negative, and once again unphysical.  The fit
illustrated in Figure \ref{chi22} employs the QCD continuum model
alone.  Hence there is no evidence of any overlap of $\chi_2$ with the
ground state nucleon in this correlation function.  While the results
illustrated here are for our intermediate value of quark mass
considered on the lattice, similar results are seen for the lighter
and heavier quark masses.  The fit parameters are summarized in Table
\ref{Tab:chi22}.  The QCD continuum threshold is not too different
{}from that for $\chi_1 \overline \chi_1$.

   Figure \ref{chi22c6} illustrates the quark mass dependence of
$\xi$.  As anticipated, $\xi$ is independent of $m_q$.  This quark
mass independence confirms the negative sign of the $m_q$ correction
appearing in the OPE for $\chi_2 \overline \chi_2$ and the use of
mean-field improved operators.  It also confirms the perturbative role
of the quark mass operator.  While the heaviest current quark mass
used in this investigation is similar to that of the strange quark, it
is still light on the scale set by the nucleon mass.

   It is also interesting to note that $\xi$ is much closer to 1 than
for the correlators of $\chi_1 \overline \chi_1$.  That this might be
the case is eluded to by the opposite signs of the leading terms of
the OPE in (\ref{OPE22}), providing the possibility of cancelations
in the short-time perturbative regime of the correlation function.

\subsection{\pmb{$\bigm < \chi_1 \overline \chi_2 +
                          \chi_2 \overline \chi_1 \bigm >$}
            Correlation Function }

   Since the square of $\lambda_2$ is small, one might be able to
recover a signal for the overlap of the nucleon ground state and
$\chi_2$ by considering the correlation function for
$\chi_1 \overline \chi_2$.  Figure \ref{chi12} illustrates a three
parameter fit from $t = 6 \to 12$ using the pole plus QCD continuum
model derived from the OPE for
${1 \over 2} ( \chi_1 \overline \chi_2 + \chi_2 \overline \chi_1)$
\begin{eqnarray}
G_2(t) &=& {3 \cdot 5 \over 2^8 \pi^4} \, \biggl ( \,
{1 \over t^6} - {4 \over 5} \, {m_q a \over t^5}
              - {2 \over 5} \, {m_q^2 a^2 \over t^4} \nonumber \\
     &&+ {8 \pi^2 \over 15} \, {\bigm < \overline q q \bigm > a^3
                \over t^3}
              + \cdots \biggr ) \, .
\label{OPE12}
\end{eqnarray}
In this fit, the nucleon ground state pole position has been fixed at
the previously determined nucleon masses.  The fit parameters are
summarized in Table \ref{Tab:chi12}.  While there is sufficient
information in the correlation function to determine a value for the
nucleon mass, the corresponding uncertainties are large.

   The leading terms of the OPEs for $\chi_2 \overline \chi_2$ and
${1 \over 2} ( \chi_1 \overline \chi_2 + \chi_2 \overline \chi_1)$
given in (\ref{OPE22}) and (\ref{OPE12}) are equivalent up to a
normalization factor of five.  Since the continuum model is
constructed to accommodate these leading terms, one expects the QCD
continuum model parameters, $s_0$ and $\xi$, for these two correlators
to be similar.  A comparison of Tables \ref{Tab:chi22} and
\ref{Tab:chi12} indicates that this is indeed the case.

   Figure \ref{chi12c6} illustrates the expected quark mass
independence of $\xi$, again confirming the negative sign of the $m_q$
correction appearing in the OPE for
${1 \over 2} ( \chi_1 \overline \chi_2 + \chi_2 \overline \chi_1)$
and the use of mean-field improved operators.  Similarly $\xi \sim 1$
as anticipated by the opposite signs of the leading terms of the OPE
of (\ref{OPE12}).

   The linear extrapolation of $\left ( \lambda_1 \lambda_2
\right)^{1/2}$ to $\kappa_{\rm cr}$ is illustrated in Figure
\ref{chi12lambda}.  The combination $\left ( \lambda_1 \lambda_2
\right)^{1/2}$ shows little sensitivity to the quark mass.  This
contrasts the dependence of $\lambda_1$ illustrated in Figure
\ref{lambda1}, where $\lambda_1$ decreases as the quarks become
lighter.  Thus, $\lambda_2$ increases for decreasing quark mass.  This
reflects the fact that $\chi_2$ vanishes in a nonrelativistic
reduction.  At the chiral limit $\left ( \lambda_1 \lambda_2
\right)^{1/2} = 0.0014(10)\ {\rm GeV}^3$.  Systematic uncertainty in
the extrapolated value of $\left ( \lambda_1 \lambda_2 \right)^{1/2}$
may be estimated using the quark mass dependence suggested by Chiral
Perturbation Theory \cite{lee94} as in \cite{leinweber94a}.  Here the
systematic uncertainty in extrapolating is negligible relative to the
statistical uncertainties.

   With the previous result $\lambda_1 = 0.013(2)\ {\rm GeV}^3$, the
nucleon coupling strength for $\chi_2$ is found to be $\lambda_2 =
0.00016(22)\ {\rm GeV}^3$, approximately 100 times smaller than
$\lambda_1$.  In short, there is only one nucleon interpolating field
that has significant overlap with the nucleon ground state, namely
\begin{equation}
\chi_1(x) = \epsilon^{abc}
                 \left ( u^{Ta}(x) C \gamma_5 d^b(x) \right ) u^c(x)
\, ,
\end{equation}
and
\begin{eqnarray}
&\lambda_{\rm SR} &\simeq \lambda_{\rm A} \simeq 2 \lambda_1 =
                          0.026(4)\ {\rm GeV}^3 \, , \\
&\lambda_2 &\simeq 0 \, .
\end{eqnarray}
The results for all the considered interpolating fields are summarized
in Table \ref{InterSum} \cite{DiffRenorm}.

\section{COMPARISON WITH OTHER CALCULATIONS}
\label{compare}

   It is worth noting that the nonperturbative QCD Sum Rule
predictions for $\lambda_{\rm SR}$ have remained quite stable over the
years, despite the fact that the early calculations have a number of
numerical errors in Wilson coefficients and anomalous dimensions
\cite{review}.  This cannot be said for the more model dependent
predictions. It is ironic that, in some cases, the model calculations
were pursued due to reservations about the reliability or validity of
the QCD Sum Rule approach.  Table \ref{CompareSR} summarizes a
collection of predictions taken and updated from Ref.\
\cite{brodsky84} and \cite{gavela89}.

   QCD Sum Rule predictions of $\lambda_2$ or $\lambda_{\rm A}$ are
more uncertain.  This is largely due to a lack of rigor in the
analysis of the sum rules.  Often, the region of validity in Borel
space is simply postulated with little regard to OPE convergence or
the size of continuum model contributions.  Many authors have fixed
the continuum threshold to a preferred value or excitation energy
rather than leaving it as a search parameter to be optimized.  The
upper limit of the Borel region must be monitored as it is a function
of the three required fit parameters, $M_N$, $\lambda_N$ and $s_0$ and
varies for different interpolating fields.  The failure to monitor
these issues in existing analyses, is largely responsible for the
apparent inconsistencies between sum rules derived from different
interpolating fields.

   Ref.\ \cite{leinweber90} is one of the few sum rule analyses
where these issues are rigorously implemented.  However, the
interpolating field $\chi_2$ was not considered there.  The sign of
the quark condensate term in (\ref{OPE22}) indicates the two sum rules
will be saturated by both positive and negative parity states.  A
careful analysis of these sum rules has not yet been attempted.

   In the QCD-SR discussion of Ref.\ \cite{ioffe81} it was concluded
that the overlap of $\chi_{\rm A}$ with the nucleon, $\lambda_{\rm
A}$, must be negligible, due to the vanishing of most of the Wilson
coefficients to dimension 8.  However, this conclusion need not be the
case.  Higher order terms of the OPE starting at dimension 9 are not
zero and could easily give rise to large overlap with the nucleon as
discovered here.  The only conclusion that may be drawn from these sum
rules is that the pole contribution is small {\it relative} to the QCD
continuum model contribution.  The pole contribution is not
necessarily small in {\it absolute} terms.

   The first five entries of Table \ref{CompareSR} summarize results
for $\lambda_{\rm SR}$ obtained from the consideration of two-point
correlation functions, and these compare favorably.  The same cannot
be said for $\lambda_{\rm A}$.  The RILM prediction \cite{schafer94a}
is $\lambda_{\rm A} = 0.040(2)$ GeV${}^3$ and is large compared to the
lattice prediction of $\lambda_{\rm A} = 0.027(5)$ GeV${}^3$.  Figure
3 of Ref.\ \cite{schafer94a} displays significant discrepancies
between a global fit of the six nucleon correlators considered and two
of the correlators.  These two correlators are both dependent on
$\lambda_{\rm A}$ and these discrepancies are not reflected in their
quoted uncertainty of $\pm 0.002$ GeV${}^3$.  Their conclusion ``What
is even more important, the simple `nucleon pole plus continuum' model
gives a very good simultaneous description for the complete set of
correlation functions'' is difficult to justify in the RILM,
particularly in light of these new results.

\section{CONCLUSIONS}
\label{conclusions}

\subsection{Interpolating Field Invariance}

   Ground state nucleon properties {\it are} independent of the
interpolating field used to excite the baryon from the vacuum.  This
invariance is satisfied in a trivial manner.  The interpolating field
which vanishes in the nonrelativistic limit,
\begin{equation}
\chi_2(x) = \epsilon^{abc}
                 \left ( u^{Ta}(x) C d^b(x) \right ) \gamma_5 u^c(x)
\, ,
\end{equation}
has negligible overlap with the nucleon ground state.  Inclusion of
$\chi_2$ components in interpolating fields only increases the
statistical uncertainties of lattice QCD correlation functions.

\subsection{Optimal Interpolator for QCD Sum Rules}

   This analysis indicates that, to a good approximation, $\chi_2$
excites pure QCD continuum.  Since $\chi_2$ has negligible overlap
with the ground state nucleon, it is tempting to simply conclude that
the optimum interpolating field is $\chi_1$.  While this is certainly
the case for lattice QCD investigations, it is not obviously the case
for QCD Sum Rule analyses.

   The optimal nucleon interpolator must involve $\chi_1$ as this
interpolator is required to maintain overlap with the ground state.
The task is to determine the optimal mixing of $\chi_2$.  The Borel
improved QCD Sum Rules for the generalized interpolator
\begin{equation}
\chi_{\cal O} = \chi_1 + \beta \chi_2 \, ,
\end{equation}
are
\begin{mathletters}
\begin{eqnarray}
&&{5 + 2 \beta + 5 \beta^2 \over 64} \, M^6 \, L^{-4/9}
\left [ 1 - e^{-s_0^2/M^2}
\left ( {s_0^4 \over 2 M^4} + {s_0^2 \over M^2} + 1 \right ) \right ]
\nonumber \\
&&\quad +{5 + 2 \beta + 5 \beta^2 \over 256} \, b \, M^2 \, L^{-4/9}
\left [ 1 - e^{-s_0^2/M^2} \right ]
\nonumber \\
&&\quad +{7 - 2 \beta - 5 \beta^2 \over 24} \, a^2 \, L^{4/9}
        -{13 - 2 \beta - 11 \beta^2 \over 96} \,
         { m_0^2 \, a^2 \over M^2}
\nonumber \\
&&\qquad = \widetilde \lambda_{\cal O}^2 \, e^{-M_N^2/M^2}
+ \widetilde \lambda_{\cal O^*}^2 \, e^{-M_{N^*}^2/M^2} \, ,
\label{QCDSR1}
\end{eqnarray}
and
\begin{eqnarray}
&&{7 - 2 \beta  - 5 \beta^2 \over 16} \, a \, M^4
\left [ 1 - e^{-s_0^2/M^2}
\left ( {s_0^2 \over M^2} + 1 \right ) \right ] \nonumber \\
&&\quad -{3 (1 - \beta^2) \over 16} \, m_0^2 \, a \, M^2 \, L^{-4/9}
\left [ 1 - e^{-s_0^2/M^2} \right ] \nonumber \\
&&\quad +{3 + 2 \beta - 5 \beta^2 \over 2^7} \, a \, b
\nonumber \\
&&\qquad = \widetilde \lambda_{\cal O}^2 \, M_N \, e^{-M_N^2/M^2}
-\widetilde \lambda_{\cal O^*}^2 \, M_{N^*} \, e^{-M_{N^*}^2/M^2} \, ,
\label{QCDSR2}
\end{eqnarray}
\label{QCDSR}
\end{mathletters}
where
\begin{mathletters}
\begin{eqnarray}
a &=& - (2 \pi)^2 \bigm < \overline q q \bigm >
   = 0.450\ {\rm GeV}^3  \, , \\
b &=&   (2 \pi)^2 \bigm < {\alpha_s \over \pi} \,
        G_{\mu \nu}^a G^{a \mu \nu} \bigm >
   = 0.474\ {\rm GeV}^4 \, , \\
m_0^2 &=& - {\bigm < \overline q \, g \, \sigma \cdot G \, q \bigm >
           \over \bigm < \overline q q \bigm >}
       = 0.65 \, , \\
L &=& {\log(M/\Lambda_{\rm QCD}) \over
       \log(\mu/\Lambda_{\rm QCD})} \, , \\
\widetilde \lambda_{\cal O} &=& (2 \pi)^2 \lambda_{\cal O} \, .
\end{eqnarray}
\end{mathletters}
Here $M$ is the Borel parameter and plays a role similar to the
inverse Euclidean time of the Lattice approach.  The condensate values
are taken from Ref.\ \cite{leinweber90} where $\mu = 0.5$ GeV and
$\Lambda_{\rm QCD} = 0.1$ GeV.  The continuum model contributions are
indicated on the left hand side of the sum rules where they appear in
brackets as subtractions from the terms of the OPE surviving in the
limit $M \to \infty$.  To aid the following discussion, both a
positive parity ground state and a negative parity excited state are
included on the right-hand-side of the sum rules.

   The first QCD-SR of (\ref{QCDSR1}) is known to have uncontrollably
large perturbative corrections to the Wilson coefficient of the
identity operator \cite{jamin88}.  In leading order, these corrections
are independent of $\beta$ and are approximately 50\%.  As a result
this sum rule must be discarded \cite{fourquark}.

   In the QCD-SR approach, approximations are made at both the quark
level and the phenomenological level.  At the quark level, the OPE is
truncated and the Wilson coefficients are calculated perturbatively.
This sets a lower limit for the Borel mass.  At the phenomenological
level, the spectral density is approximated by a pole plus the QCD
continuum model.  Maintaining ground state dominance on the
phenomenological side of the sum rule sets an upper limit on the Borel
mass.  By including $\chi_2$ components in an interpolating field, one
can reduce the continuum contributions excited by $\chi_1$ and allow a
broader Borel analysis window.

   One of the most difficult things to monitor in the QCD-SR approach
is whether the OPE is sufficiently convergent for a particular value
of Borel mass.  The lattice results presented here indicate the
$\chi_2 \overline \chi_2$ correlator has the fastest converging OPE,
as its overlap with the nucleon ground state is negligible.
Similarly, the combination $\chi_1 \overline \chi_1$ produces an OPE
with the slowest convergence, as this correlation function is
dominated by the ground state nucleon for small Borel masses.

   Hence, errors made in truncating the OPE are dominated by errors in
the $\chi_1 \overline \chi_1$ component of the general correlator.
The relative error in the OPE truncation can be reduced by adding
$\chi_2$ components to the correlator.  However, the $\chi_2$
components in the OPE are simply subtracted off again by the continuum
model terms.  Hence the relevant error is the absolute error.  For
$|\beta|
\mathrel{\raise.3ex\hbox{$<$\kern-.75em\lower1ex\hbox{$\sim$}}} 1$,
this error is dominated by $\chi_1 \overline \chi_1$ components of the
correlator.  As a result, OPE truncation errors are approximately
independent of $\beta$.  This crucial point has been neglected in
previous arguments regarding the optimal nucleon interpolating field.

   Since $\chi_2$ has negligible overlap with the nucleon, the ground
state contribution is also independent of $\beta$.  Hence, the size of
the continuum model contributions is the predominant factor in
determining the optimal interpolator.  Figure \ref{optcon} illustrates
the contributions of the continuum model terms in (\ref{QCDSR2}) for
$M=0.938$ GeV and $s_0 = 1.4$ GeV.  The following discussion is not
dependent on the precise values of these parameters.  The first point
to be made is that contributions from the continuum model are largest
for $\beta \sim -0.2$.  This selection of mixing is the worst possible
choice for extracting information on the ground state nucleon.

   Figure \ref{optcon} also indicates it is possible to have vanishing
continuum model contributions at $\beta \simeq -1.5$ or $\beta = 1$.
However, we are relying on the continuum model to account for strength
in the correlator that does not have its origin in the ground state.
Without a continuum model, one would need to include additional poles
on the right-hand-side of (\ref{QCDSR2}) to account for positive
and negative parity excitation strength.  For $\beta < -1.5$ or $\beta
> 1.0$ the correlator is negative indicating the sum rule is saturated
by a negative parity state.

   Thus the optimal interpolator is $\beta \sim -1.2$ or $\beta \sim
0.8$.  To discriminate between these two regimes, we turn to the
higher dimension operators (HDO) which do not contribute to the
continuum model.  It is these terms that provide crucial information
on whether the strength in the correlator lies in the ground state or
the excited states.  If these terms are absent, the optimal fit of the
correlator is obtained when $\widetilde\lambda_{\cal O} \to 0$ and
$s_0 \to 0$. In this case the continuum model becomes the Laplace
transform and the fit is perfect.  Hence the HDO terms should be large
in magnitude.  A change in sign from the leading terms of the OPE will
also assist in distinguishing ground state strength from excited state
strength as the change in the curvature of the correlator will be more
prominent.

   The last term of (\ref{QCDSR2}) is a HDO term, and its value is
plotted as a function of beta in Figure \ref{optcon}.  The HDO
contributions are larger for $\beta \sim -1.2$ than for $\beta \sim
0.8$.  In addition, the sign of the HDO contribution is opposite that
of the continuum model contributions.  Hence the preferred regime is
$\beta \sim -1.2$.  In fact, optimization of the three fit parameters,
$M_N$, $\widetilde\lambda_{\cal O}$ and $s_0$
$(\widetilde\lambda_{\cal O^*} = 0)$ of (\ref{QCDSR2}) for $\beta =
0.6 \to 0.8$ results in fit parameters describing pure continuum with
$s_0 \sim M_N$.  Information to separate the ground state pole from
the continuum is insufficient for $\beta = 0.6 \to 0.8$.

   In summary, the lattice results indicate OPE convergence, and
ground state pole contributions are approximately independent of
$\beta$.  Considerations of the size of continuum model contributions
and the sign and magnitude of HDO operators leads to the preferred
value of
\begin{equation}
\beta = -1.2 \pm 0.1 \, .
\end{equation}
A more precise determination of $\beta$ will depend on the details
of limits for continuum model contributions, HDO values, condensate
values, and other parameters of the sum rules.

   Hence, this analysis supports the selection of $\beta = -1$
\cite{ioffe81,ioffe83,leinweber90,furnstahl92,jin94}, over $\beta =
-0.2$ \cite{chung82,chung84,dosch89a}.  At $\beta = -0.2$, where the
leading terms of the OPE are stationary with respect to $\beta$
\cite{chung84,dosch89a}, the continuum contributions are maximal.
The positive value and small magnitude of the HDO indicates the
stability of the leading terms of the OPE will not be realized as
stability in the ground state mass, coupling, nor in the continuum
threshold.

\subsection{Future Investigations}

   These techniques may be used to determine the optimal interpolating
field for any sum rule involving $\chi_1$ and $\chi_2$ components.
Each sum rule will have an optimal selection for $\beta$.  The overlap
of spin-1/2 and spin-3/2 interpolating fields are known to yield
nucleon sum rules which offer stability in the fit parameters
\cite{leinweber90} that cannot be obtained from the more common sum
rules considered here.  It will be interesting to discover if the
historical selection of $\beta = -1$ is indeed optimal.

   While it is important to establish the optimal mixing of
interpolating fields for QCD-SR analyses, one should not overlook the
fact that there is a range of values for $\beta$ where the sum rules
are expected to work.  Moreover, the ground state contribution to all
these sum rules is equivalent to the 1\% level.  In other words, the
right-hand-side of (\ref{QCDSR2}) for a single pole plus continuum
model is independent of $\beta$.  After the first sum rule is written
down, additional sum rules may be introduced with merely one new fit
parameter $(s_0)$ per sum rule.  Since direct instanton contributions
to the sum rules are not independent of $\beta$ \cite{forkel93}, one
has an excellent opportunity to see if direct instanton contributions
really are necessary to maintain sum-rule consistency
\cite{leinweber94c}.

   Future lattice QCD investigations should aim to make a direct
comparison of the OPE and lattice correlation functions.  The Wilson
coefficients and vacuum expectation values of normal ordered operators
could be determined directly from OPE fits to the lattice data.  Such
a comparison would test the validity of the OPE in the nonperturbative
sector and our understanding of quantum field theory \cite{wilczek93}.

   A direct comparison of lattice and continuum formalisms requires
the use of an improved \cite{gabrielli91} or perfect
\cite{hasenfratz93} lattice action to reduce or eliminate lattice
anisotropy in the short time regime of lattice correlation
functions.  Alternatively, the Wilson coefficients of the
Euclidean-space correlation function may be derived via lattice
perturbation theory.

   An extremely fine lattice spacing is required to provide a
sufficient number of lattice sites within the radius of convergence of
the OPE.  In the most optimistic case, the two invariant nucleon sum
rules of a given interpolator could be isolated such that, to
dimension eight, each correlation function would have up to four
parameters to be determined when extracting OPE coefficients.
Ultraviolet cutoff considerations, OPE convergence issues and the need
for error estimates in the fit parameters place the lattice spacing at
less than 0.05~fm.

   Essential information on the importance of direct instanton
contributions to the OPE can be obtained from such an investigation.
The OPE coefficients and vacuum expectation values are determined
first by matching OPE and lattice correlation functions in which there
are no direct instanton contributions.  This approach determines the
OPE in a self consistent manner.  Then, other correlators in which
direct instanton contributions are argued to be important
\cite{forkel93} can be examined.  Discrepancies between the OPE
and lattice correlation functions would signal the possible importance
of direct instanton contributions.  If existing predictions for direct
instanton contributions to the OPE resolve the differences in the
correlators, then one has compelling evidence of a non-trivial role
for direct instanton contributions to the OPE.  The importance of such
investigations warrants further effort in this direction.

\acknowledgements

   The correlation functions used in this analysis were obtained in
collaboration with Terry Draper and Richard Woloshyn in Ref.\
\cite{leinweber91}.  I wish to thank Dick Furnstahl and Xuemin Jin for
a number of interesting and helpful discussions.  Thanks also to
Marina Nielsen for her contributions and discussions surrounding the
generalized sum rules of (\ref{QCDSR}).  This research is supported by
the National Science Foundation under grants PHY-9203145, PHY-9258270,
PHY-9207889 and PHY-9102922.

%\bibliography{c:/biblio/biblio,altinter}
%\bibliography{/home/derek/biblio/biblio,AltInter}

\begin{thebibliography}{10}
\bibitem{correlations} This proved sufficient to achieve good
statistical independence of the ensemble. An examination of a number
of observables, configuration by configuration, revealed no apparent
long-range correlation in our gauge configuration sample.
\bibitem{DiffRenorm} The values reported here differ from that
presented in Ref.\ \protect\cite{leinweber93c} due to minor
differences in the renormalization procedure.
\end{thebibliography}
% below {nh} for North-Holland, {aip} for AIP journals
%\bibliographystyle{prsty}

\mediumtext
\begin{table}
\caption{$\bigm < \chi_1 \overline\chi_1 \bigm >$:
         Four parameter search for the pole plus QCD continuum model.}
\label{Tab:chi11C5ope}
\setdec 0.000000
\begin{tabular}{lcccc}
Parameter  &$\kappa_1=0.152$ &$\kappa_2=0.154$ &$\kappa_3=0.156$
           &$\kappa_{\rm cr}=0.159\,8(2)$ \\
\tableline
$M_N a$
 &\dec 1.109(8)   &\dec 0.983(8)  &\dec 0.858(8)
 &\dec 0.628(17)\tablenotemark[1] \\
$\lambda_1 a^3$ $(\times 10^{-2})$
 &\dec 1.17(5)    &\dec 0.94(4)   &\dec 0.75(3)   &\dec 0.38(7)  \\
$s_0 a$
 &\dec 1.68(3)    &\dec 1.58(3)   &\dec 1.49(4)   &\dec 1.32(7)  \\
$\xi$
 &\dec 6.83(10)   &\dec 6.74(9)   &\dec 6.62(9)   &\dec 6.42(19) \\
$\xi$ from OPE fit
 &\dec 5.3(1)     &\dec 5.6(1)    &\dec 5.8(1)    & \\
\end{tabular}
\tablenotetext[1]{The physical proton mass sets the lattice spacing
$a=0.132(4)$ fm.}
\end{table}

\begin{table}
\caption{$\bigm < \chi_2 \overline\chi_2 \bigm >$:
         Two parameter search for the pure QCD continuum model.}
\label{Tab:chi22}
\setdec 0.000000
\begin{tabular}{lcccc}
Parameter  &$\kappa_1=0.152$ &$\kappa_2=0.154$ &$\kappa_3=0.156$
           &$\kappa_{\rm cr}=0.159\,8(2)$\\
\tableline
$s_0 a$
 &\dec 1.58(2)  &\dec 1.48(2) &\dec 1.40(2) &\dec 1.23(5)  \\
$\xi$
 &\dec 1.55(4)  &\dec 1.55(4) &\dec 1.56(4) &\dec 1.56(8) \\
$\xi$ from OPE fit
 &\dec 1.36(6)  &\dec 1.34(6) &\dec 1.35(6) & \\
\end{tabular}
\end{table}

\begin{table}
\caption{${1 \over 2}
          \bigm < \chi_1 \overline\chi_2 + \chi_2 \overline\chi_1
          \bigm >$:
         Three parameter search for the pole plus QCD continuum model.
$M_N$ has been fixed to the previously determined lattice values.}
\label{Tab:chi12}
\setdec 0.000000
\begin{tabular}{lcccc}
Parameter  &$\kappa_1=0.152$ &$\kappa_2=0.154$ &$\kappa_3=0.156$
           &$\kappa_{\rm cr}=0.159\,8(2)$\\
\tableline
$\left(\lambda_1\lambda_2\right)^{1/2} a^3$ $(\times 10^{-3})$
 &\dec 0.50(14)  &\dec 0.42(15)  &\dec 0.47(14)  &\dec 0.43(30)  \\
$s_0 a$
 &\dec 1.55(5)   &\dec 1.45(6)   &\dec 1.36(7)   &\dec 1.18(14)  \\
$\xi$
 &\dec 1.57(4)   &\dec 1.57(4)   &\dec 1.57(4)   &\dec 1.57(10) \\
$\xi$ from OPE fit
 &\dec 1.42(4)   &\dec 1.41(5)   &\dec 1.42(4)   & \\
\end{tabular}
\end{table}

\begin{table}
\caption{Summary of Lattice Results for the pole plus QCD continuum
         model.}
\label{InterSum}
\setdec 0.00(00)
\begin{tabular}{lcccc}
Interpolating Fields &$M_N$ &$\lambda_N$
                     &$\xi$ &$s_0$ \\
                     &(GeV) &(GeV${}^3$)
                     &                    &(GeV) \\
\tableline
$\chi_1 = \epsilon^{abc}
          \left ( u^a \, C \gamma_5 \, d^b \right ) u^c$
 &\dec 0.938\tablenotemark[1]  &\dec 0.013(2)
 &\dec 6.42(19) &\dec 1.98(11)  \\
$\chi_2 = \epsilon^{abc}
          \left ( u^a \, C \, d^b \right ) \gamma_5 u^c$
 &Not seen      &\dec 0.00016(22)\tablenotemark[2]
                             &\dec 1.56(7) &\dec 1.84(7)  \\
$ {1 \over 2} \left ( \chi_1 \overline\chi_2 +
                      \chi_2 \overline \chi_1 \right )$
 &Fixed         &\dec 0.0014(10)  &\dec 1.58(9) &\dec 1.76(20)  \\
$ \chi_{\rm SR} = \epsilon^{abc} \left ( u^a \, C \gamma_\mu \, u^b
 \right ) \gamma_5 \gamma^\mu d^c$
 &\dec 0.96(3)    &\dec 0.027(5)
 &\dec 4.61(14)   &\dec 1.92(11) \\
$ \chi_{\rm A} = {1 \over 2} \epsilon^{abc}
\left ( u^{Ta} C \sigma_{\mu \nu}
u^b \right ) \sigma^{\mu \nu} \gamma_5 d^c$
 &\dec 0.91(3)    &\dec 0.022(5)
 &\dec 3.54(10)   &\dec 1.82(10)  \\
\end{tabular}
\tablenotetext[1]{Defines the lattice spacing $a$. }
\tablenotetext[2]{Inferred from $\chi_1 \overline\chi_1$ and
${1 \over 2} \left ( \chi_1 \overline\chi_2 + \chi_2 \overline \chi_1
\right )$ results. }
\end{table}

\begin{table}
\caption{Comparison of predictions for $\lambda_{\rm SR}$ for
         various approaches to QCD.}
\label{CompareSR}
\setdec 0.0(0)
\begin{tabular}{llc}
Approach             &Reference
  &$\lambda_{\rm SR}$    \\
                     &
  &($\times 10^{-2}$ GeV${}^3$) \\
\tableline
Lattice (mean-field improved)
                     &This work
  &\dec 2.7(5)  \\
Lattice (conventional renormalization)
                     &Gavela {\it et al.} \cite{gavela89}
  &\dec 2.4     \\
Lattice (coordinate space)
                     &Chu {\it et al.} \cite{chu93b}
  &\dec 2.2(4)  \\
QCD Sum Rule         &Leinweber \cite{leinweber90}
  &\dec 3.1(6)  \\
Instanton Liquid     &Schafer {\it et al.} \cite{schafer94a}

  &\dec 3.2(1)  \\
Baryon wave functions $(x^2 \to 0)$
                     &Brodsky {\it et al.} \cite{brodsky84}
  &\dec 12.     \\
Quark Model          &Thomas and McKellar \cite{thomas83}
  &\dec  8.     \\
Bethe-Salpeter amplitude
                     &Tomozawa  \cite{tomozawa81,donoghue82}

  &\dec 2.5     \\
Quark Model          &Milosevic {\it et al.} \cite{milosevic82}

  &\dec  2.     \\
MIT Bag Model        &Donoghue and Golowich \cite{donoghue82}

  &\dec 1.27    \\
\end{tabular}
\end{table}

%%%%%%%%%%%%%%%%%%%%%%%%%%%%%%%%%%%%%%%%%%%%%%%%%%%%%%%%%%%%%%%%%%%%%%
%
%  Figures begin here
%
%%%%%%%%%%%%%%%%%%%%%%%%%%%%%%%%%%%%%%%%%%%%%%%%%%%%%%%%%%%%%%%%%%%%%%
\narrowtext

\begin{figure}
\caption{
The two-point correlator at $\kappa=0.154$ for the nucleon
interpolating fields $\chi_1 \overline \chi_1$ of
(\protect\ref{chiN1}).  The fits for the 13 analysis intervals are
illustrated.  The source position is at $t_0 = 4$.  Neither the source
nor $t = 5$ are included in the fit.  }
\label{NuclCorrFn}
\end{figure}

\begin{figure}
\caption{
The quark mass dependence of the nucleon coupling strength $\lambda_1$
corresponding to the interpolating field of (\protect\ref{chiN1}).}
\label{lambda1}
\end{figure}

\begin{figure}
\caption{
Lattice correlation function for interpolating fields $\chi_2
\overline \chi_2$.  The illustrated fit employs the QCD continuum
model alone. }
\label{chi22}
\end{figure}

\begin{figure}
\caption{
The dependence of $\xi$ on the quark mass.  The displayed independence
confirms the sign and magnitude for the Wilson Coefficient of the
$m_q$ term in the OPE of (\protect\ref{OPE22}) for $\chi_2 \overline
\chi_2$. }
\label{chi22c6}
\end{figure}

\begin{figure}
\caption{
Lattice correlation function for interpolating fields ${1 \over 2}
(\chi_1 \overline \chi_2 + \chi_2 \overline \chi_1)$.  The nucleon
mass has been fixed at previously determined values for this fit. }
\label{chi12}
\end{figure}

\begin{figure}
\caption{
The dependence of $\xi$ on the quark mass.  The displayed independence
confirms the sign and magnitude for the Wilson Coefficient of the
$m_q$ term in the OPE of (\protect\ref{OPE12}) for ${1 \over 2}
(\chi_1 \overline \chi_2 + \chi_2 \overline \chi_1)$. }
\label{chi12c6}
\end{figure}

\begin{figure}
\caption{
Linear extrapolation of the coupling strength $(\lambda_1
\lambda_2 )^{1/2}$ to $\kappa_{\rm cr}$.  The $y$-axis scale is one
tenth of that in Figure \protect\ref{lambda1}.  }
\label{chi12lambda}
\end{figure}

\begin{figure}
\caption{
Continuum model (solid curve) and higher dimension operator (HDO)
(dashed curve) contributions to the Borel improved QCD-SR of
(\protect\ref{QCDSR2}) plotted as a function of the interpolating
field mixing parameter $\beta$.}
\label{optcon}
\end{figure}

\end{document}